\documentclass[iop]{emulateapj}
\shorttitle{UV light curves in SLSN models}
\shortauthors{Tolstov et al.}

\usepackage{color}
\usepackage{enumerate}
\usepackage{amsmath}
\usepackage{textcomp}
\usepackage{longtable}
\usepackage{ulem}
\usepackage[final]{microtype}
\usepackage{scalefnt}
\usepackage{url}
\newcommand\smaller[2][0.85]{{\scalefont{#1}#2}}

\begin{document}

\title{Ultraviolet light curves of Gaia16apd in superluminous supernova models}

\author{Alexey Tolstov\altaffilmark{1}, Andrey Zhiglo\altaffilmark{1,2}, Ken'ichi Nomoto\altaffilmark{1}, Elena Sorokina\altaffilmark{3}, Alexandra Kozyreva\altaffilmark{4}, 
Sergei Blinnikov\altaffilmark{5,6,1}
}

\affil{\altaffilmark{1} Kavli Institute for the Physics and Mathematics of the Universe (WPI), The
University of Tokyo Institutes for Advanced Study, The University of Tokyo, 5-1-5 Kashiwanoha, Kashiwa, Chiba 277-8583, Japan} 

\affil{\altaffilmark{2}
NSC Kharkov Institute of Physics and Technology, 61108 Kharkov, Ukraine}

\affil{\altaffilmark{3} Sternberg Astronomical Institute, M.V.Lomonosov Moscow State University, 119234 Moscow, Russia}

\affil{\altaffilmark{4}
The Raymond and Beverly Sackler School of Physics and Astronomy, Tel Aviv University, Tel Aviv 69978, Israel}

\affil{\altaffilmark{5} Institute for Theoretical and Experimental Physics (ITEP), 117218 Moscow, Russia} 

\affil{\altaffilmark{6} All-Russia Research Institute of Automatics (VNIIA), 127055 Moscow, Russia}

\email{$^{*}$ E-mail: alexey.tolstov@ipmu.jp}

\submitted{Accepted for publication in the Astrophysical Journal Letters}
\journalinfo{Accepted for publication in the Astrophysical Journal Letters}
\slugcomment{Accepted for publication in the Astrophysical Journal Letters on 18 Jul 2017}

\begin{abstract}
\noindent

Observations of Gaia16apd revealed extremely luminous ultraviolet emission among superluminous supernovae (SLSNe). 
Using radiation hydrodynamics simulations we perform a comparison of UV light curves, color temperatures and photospheric velocities between the most popular SLSN models: pair-instability supernova, magnetar and interaction with circumstellar medium. We find that the interaction model is the most promising to explain the extreme UV luminosity of Gaia16apd. The differences in late-time UV emission and in color evolution found between the models can be used to link an observed SLSN event to the most appropriate model. Observations at UV wavelengths can be used to clarify the nature of SLSNe and more attention should be paid to them in future follow-up observations.

\end{abstract}



\section{INTRODUCTION}
\label{sec:intro}
\noindent


Recent detection of Gaia16apd (SN 2016eay) revealed its extraordinarily UV-bright emission among superluminous supernovae \citep{Kangas2017, Yan2017}. Gaia16apd was classified as a hydrogen-poor Type I SLSNe (SLSN-I) (peak bolometric luminosity $\sim 3 \times 10^{44}$ erg s$^{-1}$) in a faint dwarf galaxy at redshift $z$ = 0.102. It is extremely UV luminous, emitting 50 \% of its total luminosity in the wavelength range 1000 -- 2500 \AA. In comparison with SLSN PTF12dam \citep{Nicholl2013}, located at redshift $z$ = 0.107, Gaia16apd is about 2 -- 3 magnitudes brighter in Swift UV bands, having similar brightness in the optical. It is spectroscopically similar \citep{Kangas2017} to PTF12dam \citep{Nicholl2013}, SN 2010gx \citep{Pastorello2010} and SN 2011ke \citep{Inserra2013} in g- and the Swift bands. At maximum light, the estimated photospheric
temperature and velocity are 17,000 K and 14,000 km s$^{-1}$ respectively \citep{Yan2017}. The metallicity of the host galaxy is estimated as $Z$ = 0.18 $Z_{\odot}$ \citep{Nicholl2017}, which is comparable to other SLSNe.

Being a SLSN (brighter than $-21$ magnitude in all optical bands) Gaia16apd can be explained in several scenarios. At the moment there is no universally accepted model for SLSNe.
The most popular scenarios \citep[see e.g.][for review]{Quimby2014} include pair-instability supernova (PISN), a spinning-down millisecond magnetar, and interaction of the supernova ejecta with a surrounding extended and dense circumstellar matter (CSM).  

The bolometric light curve of Gaia16apd is easily fitted by a one zone magnetar model \citep{Inserra2013}, based on Arnett's analytical model for radioactive pumping of supernovae. The fits in the literature \citep{Kangas2017, Nicholl2017} give the ejecta mass $\sim$ 4 -- 13 $M_{\odot}$  and kinetic energy of the ejecta $E_{\rm 51,kin}$ $\sim$ 10 -- 20 ($E_{51}$ = $E/10^{51}$ erg), depending on the average opacity value. Gaia16apd was also proposed to be a jet-induced core collapse supernova in a negative jet feedback mechanism, where rapidly rotating neutron stars are likely to be formed \citep{Soker2017}. But all these magnetar models are oversimplified and they do not estimate the UV luminosity. 

The shock interaction mechanism can also be applied for SLSN-I  \citep{Chatzopoulos2012, Chatzopoulos2013, Sorokina2016}. This approach used in simulations of SLSN-I PTF12dam \citep{Tolstov2017} shows an excess of UV emission in the model of pulsational pair-instability supernova (PPISN) with kinetic energy of the ejecta $E_{\rm 51,kin}$ $\sim$ 10 -- 20, having a massive progentor $M_{\rm ZAMS}=100$ $M_{\odot}$. Due to an excess of UV emission, interaction models look promising to be compared with Gaia16apd observational data.

The peaks of the light curves of PISN models are too broad  \citep{Kasen2011,Dessart2013} to be used for explanation of Gaia16apd. In addition, calculated spectroscopic evolution for PISN models \citep{Dessart2013,Chatzopoulos2015} seems to be inconsistent with SLSNe and does not reveal extremely bright UV emission.

In this paper using multicolor radiation hydrodynamics simulations we perform the comparison of best-fit multicolor light curves of Gaia16apd in popular SLSN models: magnetar, PISN and interaction with CSM. The main purpose is to find out which model is the most promising for explanation of bright UV emission of Gaia16apd.

\section{Methods}
\label{sec:methods}
\noindent

For calculations of the light curves, we use 1D multigroup
radiation hydrodynamics numerical code STELLA \citep{Blinnikov1998,Blinnikov2000,blinnikov2006}. The explosion is initialized as a thermal bomb just above the
mass cut, producing a shock wave that propagates outward. 

For PISN and interaction model the energy deposition rate $L_{\rm  dep}$ is simply $L_{\rm dep}=E_{\rm dep}/t_{\rm dep}$ during relatively short time $t_{\rm dep} \sim$ 0.1s.

The energy deposition rate $L_{\rm dep}$ in magnetar model is \citep{Kasen2010}
\begin{equation}
L_{\rm dep}=\frac{E_m/t_m}{(1+t/t_m)^2},
\label{eqn1} 
\end{equation}
where the total spin energy $E_m$ and spin-down timescale $t_m$ is connected with pulsar spin period $P$ and its magnetic field $B$:
\begin{equation}
\label{eqn2}
E_m \approx 2 \times 10^{52} P^{-2}_{\rm ms} {\rm ergs} , 
\end{equation}
\begin{equation}
\label{eqn3}
t_m \approx 5B^{-2}_{14} P^{2}_{\rm ms} {\rm days} ,
\end{equation}
where $P_{\rm ms}$ = $P/1$ ms and $B_{14}$ = $B/10^{14}$ G.
Thus, we assume that all spindown energy is thermalized in the ejecta.

\section{SIMULATIONS}
\label{sec:sims}
\noindent
\subsection{Multicolor light curves: interaction with CSM}

First of all, we compare the observed Gaia16apd light curves with the best-fit interaction model of PPISN {\sc CSM40E20R16.5}, which we used in modeling PTF12dam \citep[][model CSM47]{Tolstov2017}. The progenitor has a main-sequence mass of 100 $M_{\odot}$ and a metallicity of $Z$ = $Z_{\odot}$/200. At the collapse, the mass of the C+O core is 43 $M_{\odot}$, surrounded by $\sim$ 40 $M_{\odot}$ He+C CSM (mass ratio He/C=6). In our simulations, solar metallicity is assumed in the CSM.
Models with low metallicity of CSM $Z$ = $Z_{\odot}$/200 produce essentially the same UV light curves near the peak \citep{Tolstov2017}. The explosion of this model with the energy of hypernova $E_{\rm 51,kin}=19$ shows an excess of UV emission in comparison with observed light curves of PTF12dam. The mass cut between the ejecta and the compact remnant is set at 3 $M_{\odot}$, so that the ejecta contains 6.1 $M_{\odot}$ of $^{56}$Ni \citep{Moriya2010}.

\begin{figure}
\includegraphics[width=80mm]{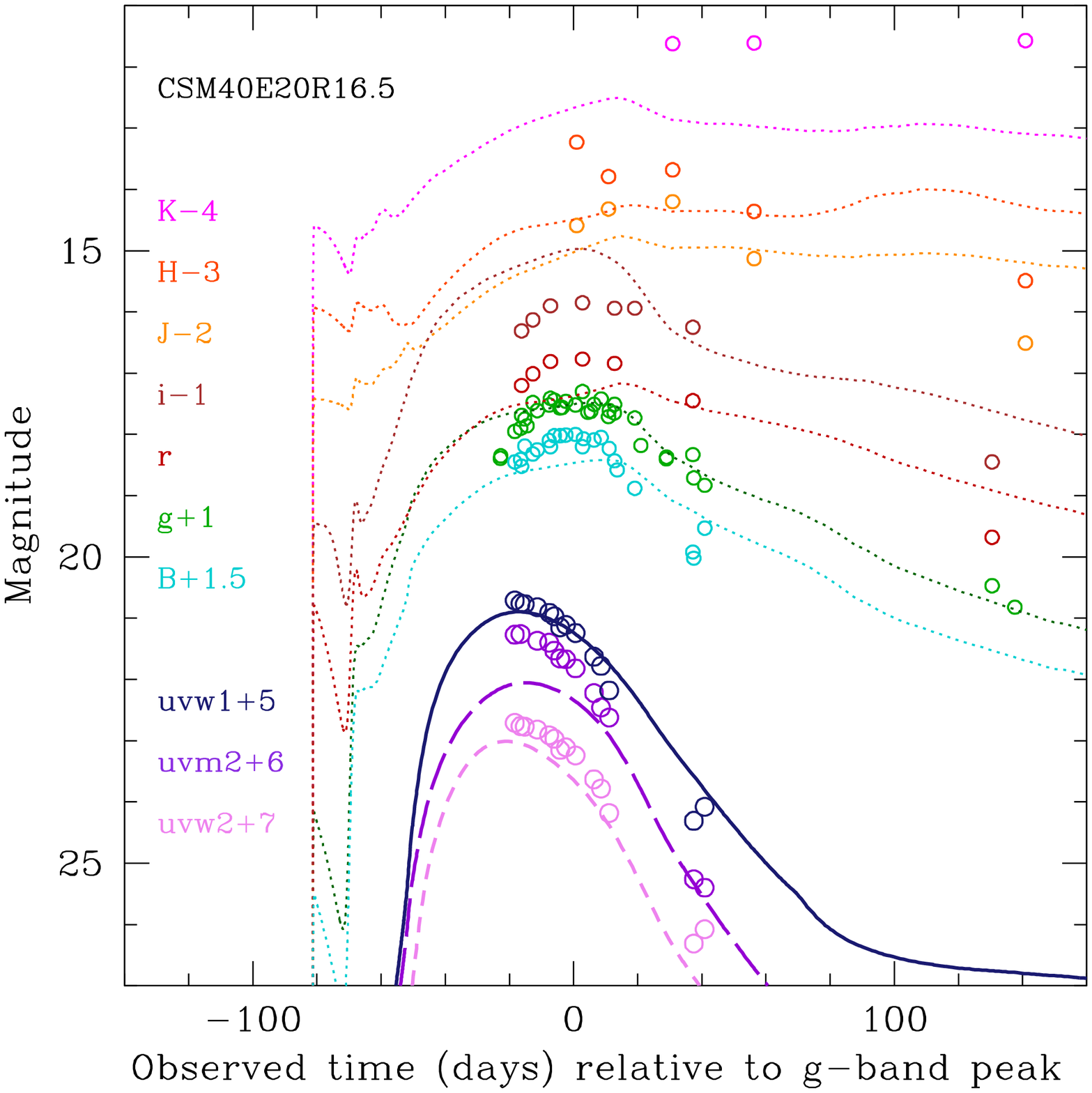}
\caption{
Multicolor light-curve simulation for Gaia16apd in the interaction model \smaller{CSM40E20R165} and comparison with observations \citep{Kangas2017}. Explosion time $t_{\rm 0} = -82$ days. UV light curves are plotted with thick lines, optical and NIR light curves are plotted with thin lines.
\\}
\label{i1}
\end{figure}

\begin{figure}
\includegraphics[width=80mm]{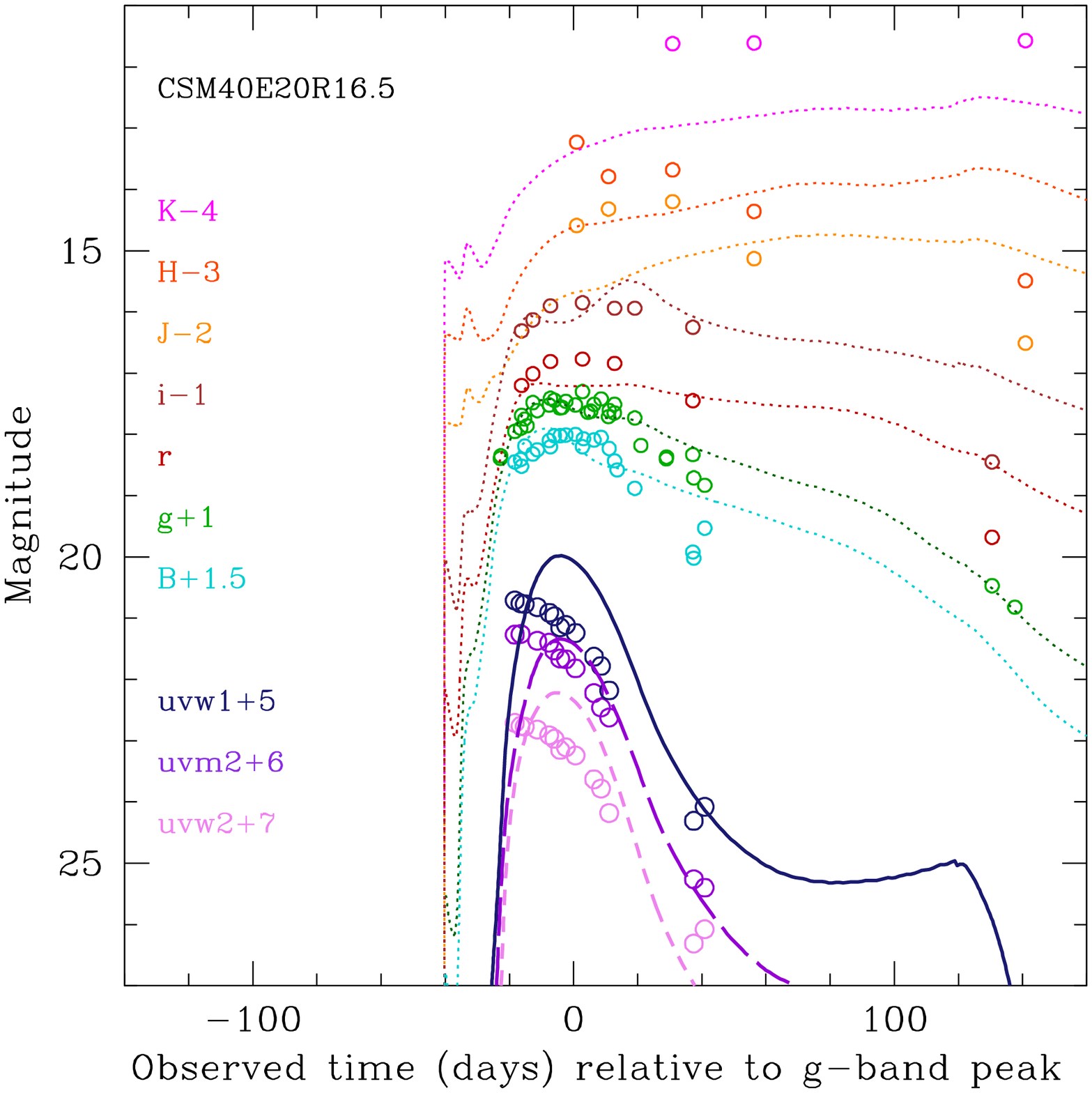}
\caption{Multicolor light-curve simulation for Gaia16apd in the interaction model {\sc
CSM20E27R162} and comparison with observations \citep{Kangas2017}. Explosion time $t_{\rm 0} = -40$ days. UV light curves are plotted with thick lines, optical and NIR light curves are plotted with thin lines.
 \\}
\label{i2}
\end{figure}

The comparison of UV light curves of Gaia16apd with the model {\sc CSM40E20R16.5} is presented in Figure \ref{i1}. Surprisingly, UV light curves have a good fit in shape of the light curves and luminosity. The raising time of the optical light curve in the model \smaller{CSM40E20R16.5} seems to be smaller to fit the observations. For this reason we consider one more interaction model with $\sim$ 20 $M_{\odot}$ C+O CSM (mass ratio O/C=4) and higher explosion energy $E_{\rm 51,kin}=27$. In C+O CSM model the raising time reduces to $t_{\rm rise}$ = 40 days, but the shape of the UV light curve becomes different (Figure \ref{i2}). Most probably the composition of CSM is not uniform mixture of C,O and He. We plan to investigate this question later by analyzing observed spectra of Gaia16apd. 

Optical light curves of interaction models have a good correspondence in luminosity with observational data and better shape for C+O CSM, while NIR requires more detailed investigation due to small number of NIR lines taken into account by standard STELLA calculation. Our calculation of \smaller{CSM40E20R16.5} model with large amount of lines (2.6 $\times$ 10$^7$) reveals the increase of peak luminosity in all bands up to 0.5 magnitude.  

Among about 100 interaction models the model {\sc CSM40E20R16.5} has the best-fit (chi-squared minimization) of UV and optical light curves to Gaia16apd. The models with high explosion energy $E_{\rm 51,kin}=30$ and $E_{\rm 51,kin}=60$ also have a good fit in UV bands, but optical peak of these models is brighter than in observations.

In our calculations, solar metallicity is assumed in the CSM.
Low metallicity affects only the tail of light curves mostly in blue and UV bands \citep[][see their
Figure 14]{Tolstov2017}. Due to lower opacity, the CSM cools down
faster and the light curve decline increases. Another
effect of lower opacity is the decrease of the radius of the
photospere, especially in UV wavelengths. The
temperature of internal CSM layers is higher, which leads to
higher luminosity at UV wavelengths.

\begin{figure}
\includegraphics[width=80mm]{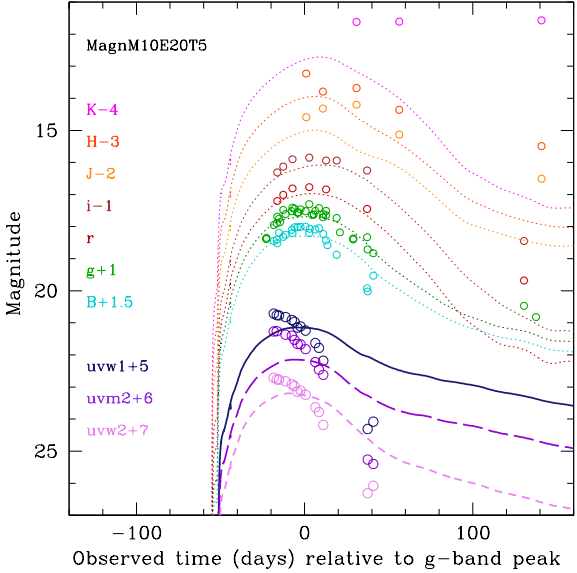}
\caption{Multicolor light-curve simulation for Gaia16apd in the magnetar model {\sc
MagnM10E20T5} and comparison with observations \citep{Kangas2017}. Explosion time $t_{\rm 0} = -54$ days. UV light curves are plotted with thick lines, optical and NIR light curves are plotted with thin lines.
\\}
\label{magn}
\end{figure}

\begin{figure}
\includegraphics[width=80mm]{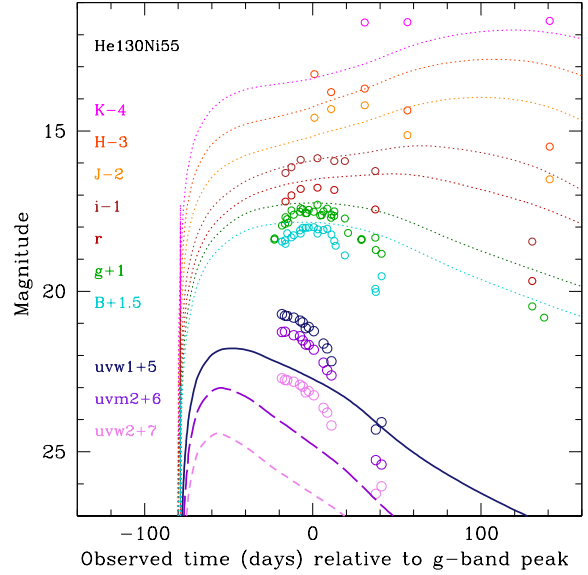}
\caption{Multicolor light-curve simulation for Gaia16apd in the PISN model {\sc
He130Ni55} and comparison with observations \citep{Kangas2017}. Explosion time $t_{\rm 0} = -81$ days. UV light curves are plotted with thick lines, optical and NIR light curves are plotted with thin lines. \\}
\label{pisn}
\end{figure}

\begin{figure}
\includegraphics[width=80mm]{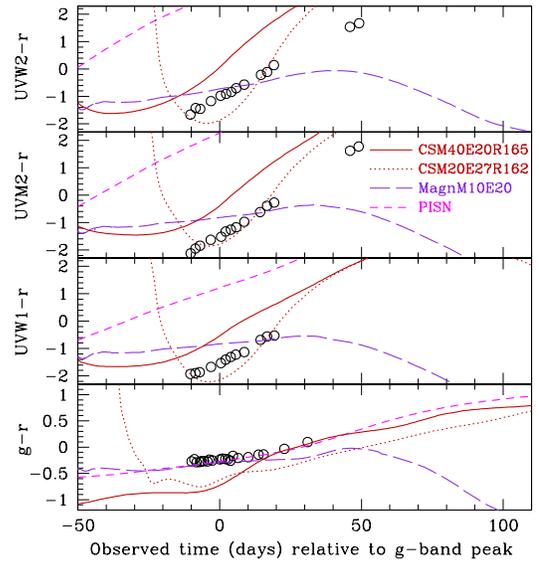}
\caption{ 
Observed color evolution of Gaia16apd \citep{Nicholl2017} compared to color evolution in interaction model {\sc CSM40E20R16.5} (solid line), interaction model {\sc CSM20E27R16.2} (dotted line), pair-instability model {\sc He130Ni55} (short dashes), and magnetar model {\sc MagnM10E20T5} (long dashes). \\ \\ }
\label{colorEvolution}
\end{figure}

\subsection{Multicolor light curves: magnetar}

For magnetar initial model 
we use the ejecta of SN 1998bw model ($M_{\rm ej}$=11 $M_{\odot}$, $M_{\rm cut}$=3 $M_{\odot}$) \citep{Iwamoto1998} at homologous expansion phase (100 s) and change the kinetic energy of the ejecta to $E_{51}$ = 1. We release the energy in the inner part of the ejecta (Eq. \ref{eqn1}) with magnetar parameters $E_m$ = $2 \times 10^{52}$ ergs, $t_m$ = $5$ days. From Eq. \ref{eqn2} and Eq. \ref{eqn3} it corresponds to $P$ = 1 ms, $B$ = $10^{14}$ G. In total we considered $\sim$ 30 models with various magnetar parameters around these values and performed chi-squared minimization to find the best-fit of UV and optical bands to observations.

The resulting light curves are presented in Figure \ref{magn}. While the optical bands have a good fit, UV and NIR luminosity is low to fit the observations. The increase of magnetar energy ($E_m > 2 \times 10^{52}$ ergs) leads to wider peak of UV light curves and it is not efficient in explaining of the shape of Gaia16apd UV light curves. Lower values of magnetar energy ($E_m < 2 \times 10^{52}$ ergs) lead to fainter UV and optical emission. Thus, magnetar parameters of the best fit of bolometric light curves \citep{Nicholl2017, Kangas2017} $P$ = 2 ms, $B$ = 2$\times$10$^{14}$ G also lead to fainter UV emission in comparison with observations.
Adding a small amount of $^{56}$Ni (up to 0.4 M$_{\odot}$) in the model increases luminosity of the light curve tails, but we did not succeed in reproducing both bright UV emission and shape of UV light curves. 

In comparison with the interaction model, the magnetar model shows differences in the evolution of the late-time UV emission. The relatively low mass ejecta of the magnetar model becomes optically thin in the UV and optical bands at $\sim$70 days after maximum light. For less energetic and less luminous magnetar models the ejecta becomes optically thin at later times and the shape of the light curves becomes close to one of the interaction model. Late-time UV emission requires more detailed simulations in optically thin regime of ejecta, where STELLA simulations are not so reliable.

\subsection{Multicolor light curves: PISN}

For the PISN scenario we choose the compact ($R$ = 4.2 $R_{\odot}$) progenitor model \smaller {He130Ni55} with the highest amount of $^{56}$Ni: $M$($^{56}$Ni)=55 $M_{\odot}$  \citep[evolutionary model of][] {HegerWoosley2002}. The total mass of the progenitor model is as small as $M$=57 $M_{\odot}$, the deposited energy $E_{\rm 51}=44$, the mass cut $M_{\rm cut}$=0.02 $M_{\odot}$. Among PISN models, the light curves of this model are characterized by the high bolometric luminosity, the relatively relatively narrow peak, and the shortest raising time. Comparison of multicolor light curves with Gaia16apd data shows that the PISN light curves are too broad to reproduce Gaia16apd and the luminosity of UV light curves are several magnitudes lower than in observations (Figure \ref{pisn}). We also checked the PISN model that was used in modeling PTF12dam \citep[][model P250]{Kozyreva2017}, but it has even wider light curve and lower UV luminosity. We thus conclude that Gaia16apd can hardly be the PISN.  

\subsection{Color evolution}

We compare the \textit{UV $-$ r} and \textit{g $-$ r} color evolution in different models (Figure \ref{colorEvolution}). The interaction model with the CO composition is in better agreement with observations in \textit{UV $-$ r} than other models. The magnetar model has a slower reddening than observations. PISN model is in good agreement with the observed reddening rate, but the model evolves about $50$ days earlier than the observed one. In contrast to \textit{UV $-$ r},
\textit{g $-$ r} color evolution is more consistent with the magnetar and the PISN model than with the interaction model.

\subsection{Color temperatures}

\begin{figure}
\includegraphics[width=80mm]{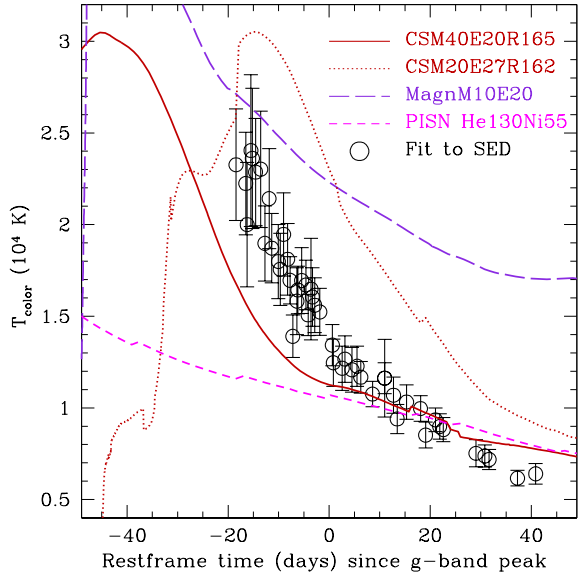}
\caption{Color temperature evolution of Gaia16apd \citep{Kangas2017}, compared with interaction model {\sc CSM40E20R16.5} (solid line), interaction model {\sc CSM20E27R16.2} (dotted line), pair-instability model {\sc He130Ni55} (short dashes), and magnetar model {\sc MagnM10E20T5} (long dashes). \\ \\}
\label{colort}
\end{figure}

We compare the evolution of the observed color temperatures $T_{\rm color}$ (temperature of the blackbody whose SED most closely fits the data) with the temperature of all described models (Figure \ref{colort}). The temperature decline rate in the interaction interaction model is a better fit to the observed values than in the magnetar and the PISN model. For more detailed comparison of spectral characteristics, the parametric study of the models is required, including variations of chemical composition.

\subsection{Photospheric Velocities}

The velocity of the photosphere (in B-band) near the optical peak in the magnetar and the PISN model $v_{ph} \sim 14,000-16,000$ km s$^{-1}$ is close to the observed $12,000-14,000$ km s$^{-1}$ \citep{Kangas2017, Yan2017}. In the best-fit interaction models the velocity is lower, $v_{ph} \sim 7,000-8,000$ km s$^{-1}$. The lower density of CSM and higher hypernova energy can increase the photospheric velocity \citep{Tolstov2017}. Also, probably, the progenitor of Gaia16apd is more compact than the progenitor that we used in interaction models. 

\section{CONCLUSIONS}
\label{sec:conclusion}

Using detailed radiation-hydrodynamics simulations,
we found a number of best-fit models in the most popular SLSN scenarios for Gaia16apd: PISN, magnetar and interaction with CSM.
The observed form of far-UV light curve in Gaia16apd and evolution of color temperatures have the best fit in interaction model of hypernova explosion. The magnetar model for the best bolometric fits has broader and less luminous peak in comparison with observations. The PISN model has too broad peak and low UV luminosity to fit the observational data of Gaia16apd. 

The differences in late-time UV emission and color evolution found between the three different models can link a particular event to the most appropriate model. The late-time UV emission for the CSM model declines more steeply as compared to that of the PISN and the magnetar spin-down model. For quantitative description of the late time emission more detailed parametric study is required including optically thin regime of ejecta. 

For the magnetar model we should note that it requires more detailed simulation of high-energy effects: pair-productions, spectral transport of gamma-rays, inverse Compton, coupling of wind and plasma. The heating of the ejecta by gamma-rays could not be as efficient as it is assumed in magnetar models. The 100\% effective thermalization of magnetar energy in the ejecta is a strong assumption. More realistic inefficient thermalization leads to lower luminosity in magnetar model \citep{Kasen2016}. Thus, the question of how efficient the conversion of spin-down energy into radiation in magnetar model is open. The preliminary more realistic simulations coupling of magnetar e$^{\pm}$-wind and plasma via pressure and energy balance and accounting for spectral transport of HEGRs show that the magnetar energy is more efficiently converted into kinetic energy, but not into radiation (Badjin 2017, in prep.). 

In our interaction model the rising time of the light curve fits to C+O composition better than to He composition. The progenitor should have a massive C+O core, but the composition of the CSM can be determined better in spectral synthesis modeling. The helium is not observed even in hydrogen-poor SLSN \citep{Yan2017ar}, but we do not exclude the presence of \lq\lq dark helium\rq\rq in the interaction model, because He features can easily be hidden in the spectra \citep{Dessart2015}. The helium ionization potential is high, but the temperature of the dense shell in the interaction model seems to be rather low to reveal helium features. Optical light curves remain almost unchanged when we add He to CO mixture in CSM while C and O are still the main absorbers \citep{Sorokina2016}.

The CSM interaction model for SLSN-I suffers in that it does not reproduce intermediate-width emission lines, like those seen routinely for superluminous Type IIn (SLSN-II, hydrogen-rich) events. Another issue is the relatively low velocity of the photosphere in our interaction model at maximum light ($\sim$ 7,000-8,000 km s$^{-1}$).  It is hard to reproduce the observed photospheric velocity of 10,000 - 15,000 km s$^{-1}$ with the CSM interaction model, but we continue to work on this topic. It seems that the problem can be solved by a suggestion that CSM is a result of rather strong preexplosion with deposited energy $E_{\rm 51,dep}=4$ \citep{Woosley2017} followed by hypernova explosion that lead to ejection of a high-velocity envelope (Blinnikov et al., in prep.).

Bright UV emission of Gaia16apd can be an on-axis emission of asymmetric explosion. The assumption of spherical symmetry adopted in our 1D approach might suffer from realistic CSM geometries (like that of Eta Carinae, or a clumpy CSM) and also from neglecting the potentially anisotropic radiative flux expected from a rapidly rotating magnetar. Hypernovae are intrinsically asymmetric and can be produced by jet-driven, bipolar explosions \citep{Maeda2002}. More realistic progenitor models and mutidimensional radiation transfer simulations will be helpful to clarify this scenario. 

To find out the nature of UV emissions of Gaia16apd and to understand how common this phenomenon is for SLSNe, both more detailed numerical simulations and future follow-up observations at UV wavelengths of SLSNe are highly in demand.
 
\acknowledgments

We thank the anonymous referee for an extensive review and useful comments.
This research is supported by the World Premier International Research Center Initiative (WPI Initiative), MEXT, Japan, and JSPS KAKENHI Grant Numbers JP16K17658, JP26400222, JP16H02168, JP17K05382. The work of ES (light curve calculation with extended line list) was supported by the Russian Scientific Foundation grant 15--12--10519. Numerical calculations were in part carried out on the CFCA cluster (XC30)
of National Astronomical Observatory of Japan.

\bibliographystyle{apj}

\begin{thebibliography}{28}
\providecommand\natexlab[1]{#1}
\providecommand\JournalTitle[1]{#1}

\bibitem[{{Blinnikov} {et~al.}(2000){Blinnikov}, {Lundqvist}, {Bartunov},
  {Nomoto}, \& {Iwamoto}}]{Blinnikov2000}
{Blinnikov}, S., {Lundqvist}, P., {Bartunov}, O., {Nomoto}, K., \& {Iwamoto},
  K. 2000, \href{http://dx.doi.org/10.1086/308588}{\JournalTitle{\apj}, 532,
  1132}

\bibitem[{{Blinnikov} {et~al.}(1998){Blinnikov}, {Eastman}, {Bartunov},
  {Popolitov}, \& {Woosley}}]{Blinnikov1998}
{Blinnikov}, S.~I., {Eastman}, R., {Bartunov}, O.~S., {Popolitov}, V.~A., \&
  {Woosley}, S.~E. 1998,
  \href{http://dx.doi.org/10.1086/305375}{\JournalTitle{\apj}, 496, 454}

\bibitem[{{Blinnikov} {et~al.}(2006){Blinnikov}, {R{\"o}pke}, {Sorokina},
  {Gieseler}, {Reinecke}, {Travaglio}, {Hillebrandt}, \&
  {Stritzinger}}]{blinnikov2006}
{Blinnikov}, S.~I., {R{\"o}pke}, F.~K., {Sorokina}, E.~I., {et~al.} 2006,
  \href{http://dx.doi.org/10.1051/0004-6361:20054594}{\JournalTitle{\aap}, 453,
  229}

\bibitem[{{Chatzopoulos} {et~al.}(2015){Chatzopoulos}, {van Rossum}, {Craig},
  {Whalen}, {Smidt}, \& {Wiggins}}]{Chatzopoulos2015}
{Chatzopoulos}, E., {van Rossum}, D.~R., {Craig}, W.~J., {et~al.} 2015,
  \href{http://dx.doi.org/10.1088/0004-637X/799/1/18}{\JournalTitle{\apj}, 799,
  18}

\bibitem[{{Chatzopoulos} \& {Wheeler}(2012)}]{Chatzopoulos2012}
{Chatzopoulos}, E., \& {Wheeler}, J.~C. 2012,
  \href{http://dx.doi.org/10.1088/0004-637X/760/2/154}{\JournalTitle{\apj},
  760, 154}

\bibitem[{{Chatzopoulos} {et~al.}(2013){Chatzopoulos}, {Wheeler}, {Vinko},
  {Horvath}, \& {Nagy}}]{Chatzopoulos2013}
{Chatzopoulos}, E., {Wheeler}, J.~C., {Vinko}, J., {Horvath}, Z.~L., \& {Nagy},
  A. 2013,
  \href{http://dx.doi.org/10.1088/0004-637X/773/1/76}{\JournalTitle{\apj}, 773,
  76}

\bibitem[{{Dessart} {et~al.}(2015){Dessart}, {Hillier}, {Woosley}, {Livne},
  {Waldman}, {Yoon}, \& {Langer}}]{Dessart2015}
{Dessart}, L., {Hillier}, D.~J., {Woosley}, S., {et~al.} 2015,
  \href{http://dx.doi.org/10.1093/mnras/stv1747}{\JournalTitle{\mnras}, 453,
  2189}

\bibitem[{{Dessart} {et~al.}(2013){Dessart}, {Waldman}, {Livne}, {Hillier}, \&
  {Blondin}}]{Dessart2013}
{Dessart}, L., {Waldman}, R., {Livne}, E., {Hillier}, D.~J., \& {Blondin}, S.
  2013, \href{http://dx.doi.org/10.1093/mnras/sts269}{\JournalTitle{\mnras},
  428, 3227}

\bibitem[{{Heger} \& {Woosley}(2002)}]{HegerWoosley2002}
{Heger}, A., \& {Woosley}, S.~E. 2002,
  \href{http://dx.doi.org/10.1086/338487}{\JournalTitle{\apj}, 567, 532}

\bibitem[{{Inserra} {et~al.}(2013){Inserra}, {Smartt}, {Jerkstrand}, {Valenti},
  {Fraser}, {Wright}, {Smith}, {Chen}, {Kotak}, {Pastorello}, {Nicholl},
  {Bresolin}, {Kudritzki}, {Benetti}, {Botticella}, {Burgett}, {Chambers},
  {Ergon}, {Flewelling}, {Fynbo}, {Geier}, {Hodapp}, {Howell}, {Huber},
  {Kaiser}, {Leloudas}, {Magill}, {Magnier}, {McCrum}, {Metcalfe}, {Price},
  {Rest}, {Sollerman}, {Sweeney}, {Taddia}, {Taubenberger}, {Tonry},
  {Wainscoat}, {Waters}, \& {Young}}]{Inserra2013}
{Inserra}, C., {Smartt}, S.~J., {Jerkstrand}, A., {et~al.} 2013,
  \href{http://dx.doi.org/10.1088/0004-637X/770/2/128}{\JournalTitle{\apj},
  770, 128}

\bibitem[{{Iwamoto} {et~al.}(1998){Iwamoto}, {Mazzali}, {Nomoto}, {Umeda},
  {Nakamura}, {Patat}, {Danziger}, {Young}, {Suzuki}, {Shigeyama},
  {Augusteijn}, {Doublier}, {Gonzalez}, {Boehnhardt}, {Brewer}, {Hainaut},
  {Lidman}, {Leibundgut}, {Cappellaro}, {Turatto}, {Galama}, {Vreeswijk},
  {Kouveliotou}, {van Paradijs}, {Pian}, {Palazzi}, \&
  {Frontera}}]{Iwamoto1998}
{Iwamoto}, K., {Mazzali}, P.~A., {Nomoto}, K., {et~al.} 1998,
  \href{http://dx.doi.org/10.1038/27155}{\JournalTitle{\nat}, 395, 672}

\bibitem[{{Kangas} {et~al.}(2017){Kangas}, {Blagorodnova}, {Mattila},
  {Lundqvist}, {Fraser}, {Burgaz}, {Cappellaro}, {Carrasco Mart{\'{\i}}nez},
  {Elias-Rosa}, {Hardy}, {Harmanen}, {Hsiao}, {Isern}, {Kankare},
  {Ko{\l}aczkowski}, {Nielsen}, {Reynolds}, {Rhodes}, {Somero}, {Stritzinger},
  \& {Wyrzykowski}}]{Kangas2017}
{Kangas}, T., {Blagorodnova}, N., {Mattila}, S., {et~al.} 2017,
  \href{http://dx.doi.org/10.1093/mnras/stx833}{\JournalTitle{\mnras}, 469,
  1246}

\bibitem[{{Kasen} \& {Bildsten}(2010)}]{Kasen2010}
{Kasen}, D., \& {Bildsten}, L. 2010,
  \href{http://dx.doi.org/10.1088/0004-637X/717/1/245}{\JournalTitle{\apj},
  717, 245}

\bibitem[{{Kasen} {et~al.}(2016){Kasen}, {Metzger}, \& {Bildsten}}]{Kasen2016}
{Kasen}, D., {Metzger}, B.~D., \& {Bildsten}, L. 2016,
  \href{http://dx.doi.org/10.3847/0004-637X/821/1/36}{\JournalTitle{\apj}, 821,
  36}

\bibitem[{{Kasen} {et~al.}(2011){Kasen}, {Woosley}, \& {Heger}}]{Kasen2011}
{Kasen}, D., {Woosley}, S.~E., \& {Heger}, A. 2011,
  \href{http://dx.doi.org/10.1088/0004-637X/734/2/102}{\JournalTitle{\apj},
  734, 102}

\bibitem[{{Kozyreva} {et~al.}(2017){Kozyreva}, {Gilmer}, {Hirschi},
  {Fr{\"o}hlich}, {Blinnikov}, {Wollaeger}, {Noebauer}, {van Rossum}, {Heger},
  {Even}, {Waldman}, {Tolstov}, {Chatzopoulos}, \& {Sorokina}}]{Kozyreva2017}
{Kozyreva}, A., {Gilmer}, M., {Hirschi}, R., {et~al.} 2017,
  \href{http://dx.doi.org/10.1093/mnras/stw2562}{\JournalTitle{\mnras}, 464,
  2854}

\bibitem[{{Maeda} {et~al.}(2002){Maeda}, {Nakamura}, {Nomoto}, {Mazzali},
  {Patat}, \& {Hachisu}}]{Maeda2002}
{Maeda}, K., {Nakamura}, T., {Nomoto}, K., {et~al.} 2002,
  \href{http://dx.doi.org/10.1086/324487}{\JournalTitle{\apj}, 565, 405}

\bibitem[{{Moriya} {et~al.}(2010){Moriya}, {Tominaga}, {Tanaka}, {Maeda}, \&
  {Nomoto}}]{Moriya2010}
{Moriya}, T., {Tominaga}, N., {Tanaka}, M., {Maeda}, K., \& {Nomoto}, K. 2010,
  \href{http://dx.doi.org/10.1088/2041-8205/717/2/L83}{\JournalTitle{\apjl},
  717, L83}

\bibitem[{{Nicholl} {et~al.}(2017){Nicholl}, {Berger}, {Margutti}, {Blanchard},
  {Milisavljevic}, {Challis}, {Metzger}, \& {Chornock}}]{Nicholl2017}
{Nicholl}, M., {Berger}, E., {Margutti}, R., {et~al.} 2017,
  \href{http://dx.doi.org/10.3847/2041-8213/aa56c5}{\JournalTitle{\apjl}, 835,
  L8}

\bibitem[{{Nicholl} {et~al.}(2013){Nicholl}, {Smartt}, {Jerkstrand}, {Inserra},
  {McCrum}, {Kotak}, {Fraser}, {Wright}, {Chen}, {Smith}, {Young}, {Sim},
  {Valenti}, {Howell}, {Bresolin}, {Kudritzki}, {Tonry}, {Huber}, {Rest},
  {Pastorello}, {Tomasella}, {Cappellaro}, {Benetti}, {Mattila}, {Kankare},
  {Kangas}, {Leloudas}, {Sollerman}, {Taddia}, {Berger}, {Chornock}, {Narayan},
  {Stubbs}, {Foley}, {Lunnan}, {Soderberg}, {Sanders}, {Milisavljevic},
  {Margutti}, {Kirshner}, {Elias-Rosa}, {Morales-Garoffolo}, {Taubenberger},
  {Botticella}, {Gezari}, {Urata}, {Rodney}, {Riess}, {Scolnic}, {Wood-Vasey},
  {Burgett}, {Chambers}, {Flewelling}, {Magnier}, {Kaiser}, {Metcalfe},
  {Morgan}, {Price}, {Sweeney}, \& {Waters}}]{Nicholl2013}
{Nicholl}, M., {Smartt}, S.~J., {Jerkstrand}, A., {et~al.} 2013,
  \href{http://dx.doi.org/10.1038/nature12569}{\JournalTitle{\nat}, 502, 346}

\bibitem[{{Pastorello} {et~al.}(2010){Pastorello}, {Smartt}, {Botticella},
  {Maguire}, {Fraser}, {Smith}, {Kotak}, {Magill}, {Valenti}, {Young},
  {Gezari}, {Bresolin}, {Kudritzki}, {Howell}, {Rest}, {Metcalfe}, {Mattila},
  {Kankare}, {Huang}, {Urata}, {Burgett}, {Chambers}, {Dombeck}, {Flewelling},
  {Grav}, {Heasley}, {Hodapp}, {Kaiser}, {Luppino}, {Lupton}, {Magnier},
  {Monet}, {Morgan}, {Onaka}, {Price}, {Rhoads}, {Siegmund}, {Stubbs},
  {Sweeney}, {Tonry}, {Wainscoat}, {Waterson}, {Waters}, \&
  {Wynn-Williams}}]{Pastorello2010}
{Pastorello}, A., {Smartt}, S.~J., {Botticella}, M.~T., {et~al.} 2010,
  \href{http://dx.doi.org/10.1088/2041-8205/724/1/L16}{\JournalTitle{\apjl},
  724, L16}

\bibitem[{{Quimby}(2014)}]{Quimby2014}
{Quimby}, R.~M. 2014, \href{http://dx.doi.org/10.1017/S1743921313009253}{ \\\url{in IAU
  Symposium, Vol. 296, Supernova Environmental} \\\url{ Impacts, ed. A.~{Ray} \& R.~A. {McCray}}, 68}

\bibitem[{{Soker}(2017)}]{Soker2017}
{Soker}, N. 2017,
  \href{http://dx.doi.org/10.3847/2041-8213/aa6a10}{\JournalTitle{\apjl}, 839,
  L6}

\bibitem[{{Sorokina} {et~al.}(2016){Sorokina}, {Blinnikov}, {Nomoto}, {Quimby},
  \& {Tolstov}}]{Sorokina2016}
{Sorokina}, E., {Blinnikov}, S., {Nomoto}, K., {Quimby}, R., \& {Tolstov}, A.
  2016,
  \href{http://dx.doi.org/10.3847/0004-637X/829/1/17}{\JournalTitle{\apj}, 829,
  17}

\bibitem[{{Tolstov} {et~al.}(2017){Tolstov}, {Nomoto}, {Blinnikov}, {Sorokina},
  {Quimby}, \& {Baklanov}}]{Tolstov2017}
{Tolstov}, A., {Nomoto}, K., {Blinnikov}, S., {et~al.} 2017,
  \href{http://dx.doi.org/10.3847/1538-4357/835/2/266}{\JournalTitle{\apj},
  835, 266}

\bibitem[{{Woosley}(2017)}]{Woosley2017}
{Woosley}, S.~E. 2017,
  \href{http://dx.doi.org/10.3847/1538-4357/836/2/244}{\JournalTitle{\apj},
  836, 244}

\bibitem[{{Yan} {et~al.}(2017{\natexlab{a}}){Yan}, {Quimby}, {Gal-Yam},
  {Brown}, {Blagorodnova}, {Ofek}, {Lunnan}, {Cooke}, {Cenko}, {Jencson}, \&
  {Kasliwal}}]{Yan2017}
{Yan}, L., {Quimby}, R., {Gal-Yam}, A., {et~al.} 2017{\natexlab{a}},
  \href{http://dx.doi.org/10.3847/1538-4357/aa6b02}{\JournalTitle{\apj}, 840,
  57}

\bibitem[{{Yan} {et~al.}(2017{\natexlab{b}}){Yan}, {Lunnan}, {Perley},
  {Gal-Yam}, {Yaron}, {Roy}, {Quimby}, {Sollerman}, {Fremling}, {Leloudas},
  {Cenko}, {Vreeswijk}, {De Cia}, {Ofek}, {Kulkarni}, {Masci}, {Rebbapragada},
  \& {Wozniak}}]{Yan2017ar}
{Yan}, L., {Lunnan}, R., {Perley}, D., {et~al.} 2017{\natexlab{b}},
  \JournalTitle{ArXiv e-prints},
  \href{http://arxiv.org/abs/1704.05061}{{\sffamily arXiv:1704.05061
  [astro-ph.HE]}}

\end{thebibliography}

\end{document}